%% file: mainpaper.tex
\documentclass[conference]{IEEEtran}
\IEEEoverridecommandlockouts
\usepackage{cite}
\usepackage{amsmath,amssymb,amsfonts}
\usepackage{algorithmic}
\usepackage{graphicx}
\usepackage{textcomp}
\usepackage{xcolor}
\usepackage{float,subfig,stfloats}
\usepackage{tikz}
\usetikzlibrary{arrows, calc, positioning, decorations.markings}
\usetikzlibrary{shapes.geometric}
\tikzstyle{block} = [rectangle, draw, fill=blue!40, text centered, minimum height=2.5em, inner sep=.2em]

\tikzstyle{blockplain}=[rectangle, draw, text centered, inner sep=.5em]
\tikzstyle{line} = [draw, thick, -latex']
\tikzstyle{operator} = [draw, shape=circle, node distance=1.5cm, line width=1pt, minimum width=1em, inner sep=-0.1pt]
\tikzstyle{branch}=[fill,shape=circle,minimum size=2pt,inner sep=0pt]
\usetikzlibrary{shapes.multipart}
\tikzset{
state2/.style={
       rectangle split,
       rectangle split parts=2,
       rectangle split part fill={orange!40,green!40},
       draw=black, thick,
       minimum height=2em,
       text width=3cm,
       inner sep=.2em,
       text centered,
       }
}
\tikzset{
state3/.style={
       rectangle split,
       rectangle split parts=3,
       rectangle split part fill={purple!40,orange!40,green!40},
       draw=black, thick,
       minimum height=2em,
       text width=3cm,
       inner sep=.2em,
       text centered,
       }
}
\def\BibTeX{{\rm B\kern-.05em{\sc i\kern-.025em b}\kern-.08em
    T\kern-.1667em\lower.7ex\hbox{E}\kern-.125emX}}
\begin{document}

\title{A Data-driven Deep Learning Approach for Bitcoin Price Forecasting
}

\author{
\IEEEauthorblockN{Parth Daxesh Modi\IEEEauthorrefmark{1}, 
                Kamyar Arshi\IEEEauthorrefmark{2}\thanks{\IEEEauthorrefmark{1}\IEEEauthorrefmark{2}These two authors contributed equally}, 
                Pertami J. Kunz\IEEEauthorrefmark{3}\thanks{The work of Pertami J. Kunz is supported by the Graduate School CE within the Centre for Computational Engineering at Technische Universit\"at Darmstadt.}, 
                Abdelhak M. Zoubir\IEEEauthorrefmark{4}}
\IEEEauthorblockA{\IEEEauthorrefmark{3}\textit{Grad. School of Comp. Eng.}, 
\IEEEauthorrefmark{3}\IEEEauthorrefmark{4}\textit{Signal Processing Group} \\
\textit{
    Darmstadt University of Technology}\\
    \IEEEauthorrefmark{1}modiparth527@gmail.com, 
    \IEEEauthorrefmark{2}kamyararshi@gmail.com, 
    \IEEEauthorrefmark{3}pertami.kunz@ieee.org, 
    \IEEEauthorrefmark{4}zoubir@ieee.org}

}

\IEEEoverridecommandlockouts
\IEEEpubid{\makebox[\columnwidth]{979-8-3503-3959-8/23/\$31.00~\copyright2023 IEEE \hfill} \hspace{\columnsep}\makebox[\columnwidth]{ }}

\maketitle
	
\IEEEpubidadjcol

\begin{abstract}
Bitcoin as a cryptocurrency has been one of the most important digital coins and the first decentralized digital currency. Deep neural networks, on the other hand, has shown promising results recently; however, we require huge amount of high-quality data to leverage their power. There are some techniques such as augmentation that can help us with increasing the dataset size, but we cannot exploit them on historical bitcoin data. As a result, we propose a shallow Bidirectional-LSTM (Bi-LSTM) model, fed with feature engineered data using our proposed method to forecast bitcoin closing prices in a daily time frame. We compare the performance with that of other forecasting methods, and show that with the help of the proposed feature engineering method, a shallow deep neural network out-performs other popular price forecasting models.

\end{abstract}


\section{Introduction}\label{I}
Bitcoin is the first decentralized cryptocurrency that has become popular and widespread in the past years. It was introduced initially by an unknown identity under the pseudonym of Satoshi Nakamoto \cite{nakamoto2008bitcoin}, and it was built
 without the need for any intermediate party in making transactions, thereby making it secure by verifying each transaction in a publicly distributed ledger called the blockchain \cite{guo2018bitcoin}. Bitcoin’s transactions run 24/7,
 and the currency is exchangeable in almost all cryptocurrency exchanges. 
 Furthermore, Bitcoin allows traders and investors to benefit from better portfolio management \cite{guo2021mrc}. Despite all the upsides, the price of bitcoin has experienced 
 drastic rises and falls
 showing its high volatility and risk, hence bitcoin price prediction 
 has always been an attractive topic among traders and the research community.

Thanks to the era of big data, deep learning algorithms have been showing their dominance in different fields such as logistics, computer vision, finance, and signal processing. 
There has been a lot of research 
in previous years using machine learning methods for crypto market forecasting, and deep learning methods play a big role in most of it \cite{guo2021mrc}. One of the famous deep learning networks that are a state-of-the-art method in processing sequential data is 
the Long-Short-Term-Memory (LSTM) network 
\cite{hochreiter1997long}, which is capable of finding long-term as well as short-term hidden dependency sequential structures in data such as natural language. Since the bitcoin price also follows a sequential structure, meaning the price of each time frame depends on previous prices in the order of time, LSTM networks can be exploited to predict bitcoin’s price in a defined time proportion. There have been further studies around using time-series networks for cryptocurrency price forecasting such as \cite{dutta2020gated}, where Dutta \textit{et al.} introduced a robust feature engineering with a simpler time-series network for prediction, or Wu \textit{et al.} \cite{8637486}, where they have proposed two LSTM based models and compared the performance on the price prediction. In Jaquart \textit{et al.} \cite{jaquart2021short} various machine learning models are tested for price prediction in different time frames, ranging from one minute to 60 minutes, and it was concluded that recurrent neural networks and gradient-boosting classifiers are well-suited for such a task. In our proposed method, we have used a novel feature extraction and selection method, in which we use technical analysis indicators for the former and a Random Forest Regressor for the latter, to exploit the best possible features for bitcoin closing price forecasting in a daily time frame, and feed our features into a shallow Bi-LSTM network to not only decrease computational complexity but also having a promising performance. 

Since deep learning models require a vast amount of data, one of the main challenges in bitcoin price prediction is that the available data is limited and none of the data augmentation tricks works. Therefore, we cannot simply use as many layers in our network as we want. As a result, we propose a method in that not only useful features are exploited with the help of feature engineering, but also our model is kept shallow and not computationally heavy.

Our main contribution is in the feature engineering and selection steps as well as the shallow architecture that completes the whole pipeline of Bitcoin price prediction. In Section \ref{Methodologies} we elaborate the features used and the process of feature extraction and selection. We describe the various models we have exploited and compared them with our proposed method. Finally, we show our results and conclusion in Section \ref{ResultConc}.



\section{Methodologies}\label{Methodologies}

One of the most prominent figures used in price analysis in finance is the OCHL chart, which includes four prices for each defined time frame. \textbf{Open}, \textbf{Close}, \textbf{High}, and \textbf{Low} 
refer to, respectively, the opening price, closing price, highest price, and the lowest price of a transaction in the respective time frame. 
We used the bitcoin's OCHL prices for each day from January 2013 until September 2021 while extracting and selecting some of the most important indicators for our task as our dataset for training and validation. We utilized InvestPy API \cite{investpy} to scrap the historical bitcoin prices.

The raw transaction data 
show high correlations with one another. We aim to predict the closing price of the next day using this dataset.
Using raw transactions may lead to overfitting of the machine learning (ML) models due to the aforementioned high correlation among the features. Therefore, we proposed a feature engineering method to extract and select the best features for training, with respect to our target task.

\subsection{Proposed Feature Extraction and Selection}
\label{SSFeatExtr}
Other than collecting OCHL daily bitcoin transaction prices (4 features), we utilized Bitinfocharts\footnote{https://bitinfocharts.com/} to extract 19 raw features: transactions in blockchain,
 average block size, sent by address, average mining difficulty, average hashrate, mining profitability, sent coins in USD, average transaction fees, median transaction fees, average block time, average transaction value, median transaction value, tweets, google trends, active addresses, top 100 to total percentage,
 average fee to reward, number of coins in circulation, and miner revenue.
 
For each of these (4+19) features, 3 windows (7 days, 30 days, 90 days) of 12 technical indicators were derived: the moving average (MA), weighted MA, Exponential MA, double exponential MA, triple exponential MA, standard deviation, variance, relative strength index, rate of change, upper and lower Bolliger bands \cite{bollinger1992using}, and MA convergence divergence. 

\begin{figure*}[htbp]
\centerline{\includegraphics[width=.9\linewidth]{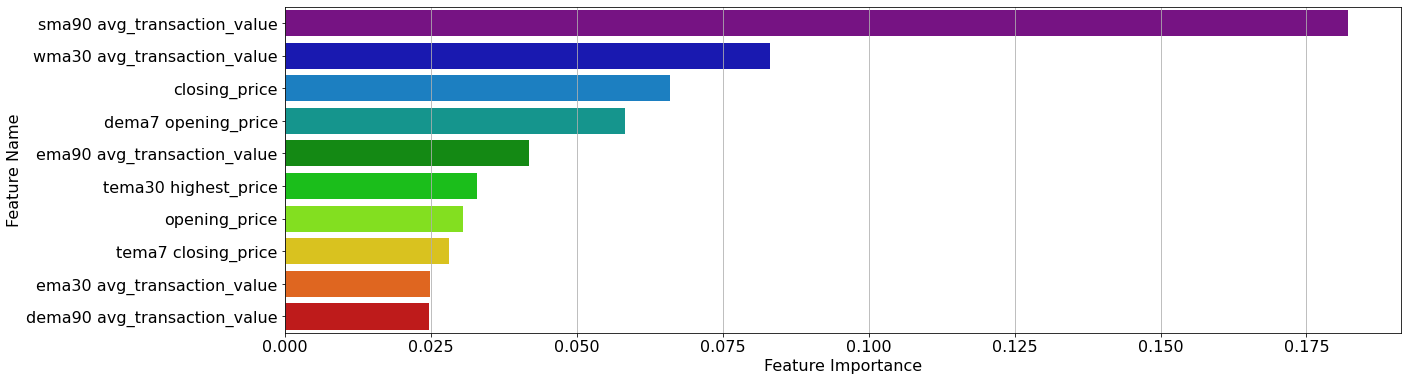}}
\caption{Top 10 most important features after feature extraction using the technical indicators and ranking them using a Random Forest Regressor, where ema=Exponential Moving Average, wma=Weighted Exponential Moving Average, dema=Double Exponential Moving Average, tema=Triple Exponential Moving Average, avg=Average, and the numbers (7, 30, 90) refer to the window sizes.
}
\label{fig:feat}
\end{figure*}

In total, we derived 23$\times$3$\times$12=828 new features. Including the raw features, we fed in total 828+23=851 features to a robust scaler, that scales the data according to the interquartile range (IQR), to make the scales of all the features the same, and also that our ML models be less affected by outliers. Subsequently, we used a Random Forest (RF) Regressor \cite{breiman2001random} to evaluate the importance of each feature given our regression task, which is predicting the closing price of the next day. From the results, we only used the top 10 most important features ranked by the RF Regressor 
(Fig.~\ref{fig:feat}).

\subsection{Train/Test Split}\label{SSTrainTest}
\begin{figure*}[htbp]
\centerline{\includegraphics[width=.9\linewidth]{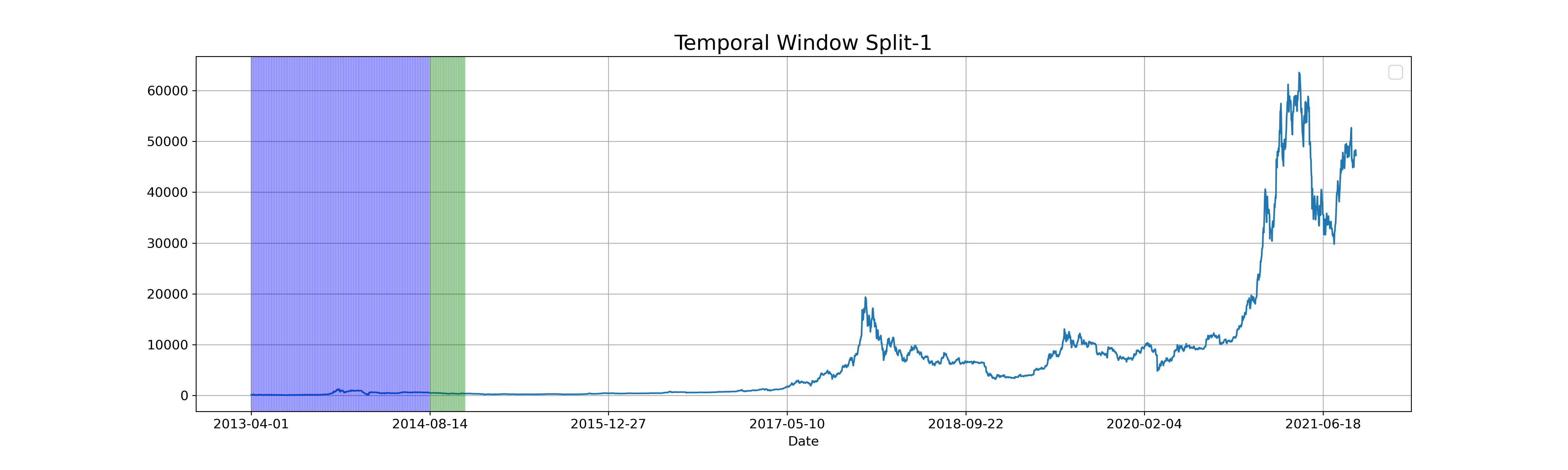}}
\caption{Train set and Test set splitting in the first batch. The blue box is the first training batch and the green box is the first test batch. This is done on sequentially on the dataset shown in this figure.}
\label{fig:split}
\end{figure*}

As the bitcoin price is highly volatile, from 100 USD in 2013 to 63K USD in 2021, it is hard to train a model that generalizes well on such a huge dynamic range. Thus, we train multiple models by data splitting so as to include different time frames with different price ranges and thus various seasonality and trends in the sequence. Each training batch split consists of 500 data points, and the next 100 data points in the sequence are used as validation (testing) batch. This process is applied for all the available data points and is illustrated in Fig.~\ref{fig:split}.

Next we explain the methods of each model we exploited, followed by a description of our proposed model’s building blocks.

\subsection{Support Vector Regressor (SVR)}
Support Vector Machines \cite{boser1992training} are one of the most powerful supervised learning algorithms. They are versatile and able to perform nonlinear and linear classification and regression. SVR works in the same way as an SVM Classifier works, but instead of finding the hyperplane that maximizes the distance of the closest data points of two different classes, it tries to fit as many data points as possible on the hyperplane while limiting margin violations \cite{10.1115/1.1897403}.
We have implemented this algorithm using the kernel Radial Basis Function (RBF), that has the benefit of being stationary and isotropic.

Another main reason of using SVR is that it works well on small datasets, and since it is indeed our case due to the splitting approach, as explained in the previous section.

\subsection{LSTM}
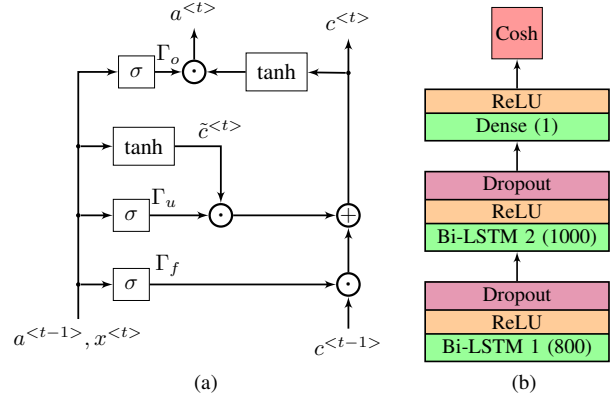
\begin{figure}
    \subfloat[]{
        \resizebox{0.6\linewidth}{!}{%
            \input{diag3}
        }
    }
    \subfloat[]{
        \resizebox{0.3\linewidth}{!}{%
            \input{diag2}
        }
    }
    \caption{(a) The individual LSTM cell and (b) the proposed Bi-LSTM neural network. Cosh is the Cosine Hyperbolic loss.}
\label{fig:ab}
\end{figure}

While using all the above models, the sequential relationship in the time series is not taken into consideration. The statistical models ARMA, and GARCH did so, but they lack in capturing the non-linearity in the time series. Furthermore, in a time-series dataset, both the long-term and short-term dependencies may be important. As a result, using simple RNN blocks might lead to gradient vanishing problems and will not consider long-term relations in the data. At this point, using Long-Short-Term-Memory neural networks will solve the aforementioned dilemma. The structure of one LSTM cell is shown in Fig.~\ref{fig:ab} and the output is calculated as follows \cite{ng2017mlyearning}:

\begin{align}
    \Tilde{c}^{<t>} &= \tanh(W_c[a^{<t-1>}, x^{<t>}]+b_c) \nonumber \\
    \Gamma_u \quad &= \sigma(W_u[a^{<t-1>}, x^{<t>}]+b_u) \nonumber \\
    \Gamma_f \quad &= \sigma(W_f[a^{<t-1>}, x^{<t>}]+b_f) \nonumber \\
    \Gamma_o \quad &= \sigma(W_o[a^{<t-1>}, x^{<t>}]+b_o)\nonumber\\
    c^{<t>} &= \Gamma_u\odot\Tilde{c^{<t>}} + \Gamma_f \odot c^{<t-1>}\nonumber \\
    a^{<t>} &= \Gamma_o\odot \tanh(c^{<t>}),
\end{align}

where $\Tilde{c}$ is the cell input activation vector, $\Gamma_u, \Gamma_f,$ and $\Gamma_o$ are the update, forget, and output gates activation vectors, respectively, $c$ and $a$ are the cell the hidden state vectors, $\sigma$ is the sigmoid function, $W$ and $b$ refer to the weight matrices and bias vector parameters, $\odot$ sign is element-wise multiplication, and $<t>$ means at time step $t$.
The architecture of the LSTM neural network we have used is exactly the same as our proposed Bi-LSTM model, and instead of the Bi-LSTM cells we have LSTM cells.
\subsection{Bidirectional-LSTM}
LSTM neural networks are fed with a sequence in the dataset in order from the beginning of the series at time 0 until the end of the sequence. However, sometimes there are hidden relations in a sequence when looking at it from the other way, meaning in the reverse descending order. To exploit this, we can use Bi-LSTM networks \cite{article}, which not only do the same thing as LSTM does, but also they take input from the last element in a sequence and continue going back to the start of it. This makes the neural network capable of finding hidden sequential relations in both ways.

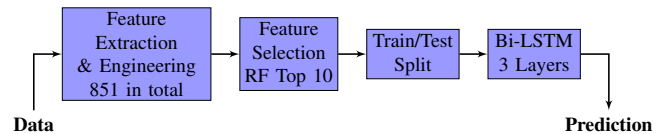
\begin{figure}
    \centering
    \resizebox{\linewidth}{!}{%
        \input{diag1}
    }
    \caption{Whole proposed pipeline. The final block, Bi-LSTM 3 Layers, contains the cells illustrated in Fig. \ref{fig:ab}(b).}
    \label{fig:diag1}
\end{figure}

\subsection{Proposed architecture}
Our proposed architecture consists of three layers. The first and the second layer are Bi-LSTM cells and each is followed by dropout layers in training to avoid overfitting. At the end, we have a single neuron fully-connected layer to output the prediction. Each layer's output goes through a ReLU \cite{DBLP:journals/corr/abs-1803-08375} activation function since the price cannot be a negative number and to avoid the vanishing gradient problem. We used a hyperbolic cosine loss as our loss function due to two main reasons: (a) It behaves stable during the gradient descent search and (b) is also not affected by sudden disparate predictions \cite{https://doi.org/10.48550/arxiv.2101.10427}. The architecture of the proposed Bi-LSTM structure is shown in Fig.~\ref{fig:ab}, and the whole proposed pipeline is illustrated in Fig. \ref{fig:diag1}.

\section{Results and Conclusion}\label{ResultConc}
Each model is trained on the transformed dataset with the selected features, and divided into training and testing batches, as discussed in Section \ref{Methodologies}. The training is performed to predict the next closing price of bitcoin. As shown in Table \ref{table:1}, the mean value of the three error types, RMSE, MAE, and MAPE, is measured for the predictions on both the training sets and test sets. Since there might be outliers while taking the mean, we also have provided the median performance metric values of each training and test batch in Table \ref{table:1}. The results of the ML models of Section \ref{Methodologies} are presented in the Table \ref{table:1} along with the results of a linear regression (LR) model to compare each model performance with a baseline.

We can observe in Table \ref{table:1} that our proposed Bi-LSTM is performing more consistently and better compared to other models. Furthermore, it is clearly presented in the table that the performance of the proposed model on the test batches has the fewest outliers, since the median and the mean MAPE are the same, $3.16\%$. Please note that we have not included the comparison with the ARMA, ARIMA and the GARCH models since they are trained on the entire sequence of data set (without splitting) and thus it is not a fair comparison to describe.

As a summary, this study has proposed a data-driven approach for predicting Bitcoin's closing price, using various methods of feature extraction, selection, and data splitting, alongside a proposed Bi-LSTM neural network architecture to tackle the high volatility and time series dependencies in bitcoin price. We have also compared and explained various time-series and ML methods with their pros and cons and clarified the reason of using neural networks and in specific Bi-LSTM networks. Eventually, we have compared the results and showed that the proposed shallow Bi-LSTM architecture performs the best and the most consistently on average. Other than being computationally optimized, this forecasting model may aid traders working with cryptocurrency, especially since the crypto market is 24/7, and with the high volatility bitcoin price has, it could be a good metric for AI-assisted trading for professional traders.


\renewcommand{\arraystretch}{1.2}
\begin{table}[htbp]
	    \setlength\tabcolsep{2.5pt}
\caption{Mean and Median Performance Comparison}
\begin{center}
\begin{tabular}{|c|c|c|c|c|c|c|}
\hline
\textbf{}&\multicolumn{6}{|c|}{\textbf{Mean of Metrics}} \\
\cline{2-7} 
\textbf{Methods} & \multicolumn{2}{|c|}{\textbf{\textit{RMSE}}}& \multicolumn{2}{|c|}{\textbf{\textit{MAE}}}& \multicolumn{2}{|c|}{\textbf{\textit{MAPE}}} \\
  & train & test & train & test & train & test \\
\hline
LR & 378.9091 & 674.032 & 246.3711 & 546.109 & 0.1945 & 0.22664\\
SVR & 380.7813& 898.2263& 239.8385 & 738.6972 & 0.1370 & 0.1850\\
LSTM & \textbf{262.8562}& 455.5994 & \textbf{149.1471}& 377.3157 & \textbf{0.0297} &  0.0337\\
Proposed & 268.3314& \textbf{450.3816} & 152.8135& \textbf{334.6625} & 0.0312 & \textbf{0.0316}\\
\cline{1-7}
&\multicolumn{6}{|c|}{\textbf{Median of Metrics}} \\
\cline{1-7}
LR & 298.8742&  373.9383 & 216.6750 & 312.1933 & 0.0554 & .0687\\
SVR & 299.1291& 403.4483 & 201.6113 & 340.4691 & 0.05493 & 0.0856\\
LSTM & \textbf{211.8146} & 215.4055 & \textbf{122.4450} & 154.995 & \textbf{0.0258} & \textbf{0.03073}\\
Proposed$^{\mathrm{1}}$& 215.9530 & \textbf{197.4914}& 125.0576 & \textbf{135.7671} & 0.02647 & 0.0316\\
\hline
\multicolumn{7}{l}{$^{\mathrm{1}}$The Proposed shallow Bi-LSTM model has the same MAPE for the mean and }\\
\multicolumn{7}{l}{median on the test set}
\end{tabular}
\label{tab1}
\end{center}
\label{table:1}
\end{table}

\section*{Acknowledgment}

We thank Abhishek Deshmukh 
 and Ekican Cetin 
 for their involvement in the earlier stage of this project.

\bibliographystyle{plain}
\bibliography{IEEEexample}
\end{document}

%% file: diag3.tex
\begin{tikzpicture}[auto]
    \node[](a0) {$a^{<t-1>}, x^{<t>}$}; 
    \node[branch, above=0.5cm of a0](b1) {};
    \node[blockplain, right=0.5cm of b1](s1) {$\sigma$};
    \node[branch, above=of b1](b2) {};
    \node[blockplain, right=0.5cm of b2](s2) {$\sigma$};
    \node[operator, right=2cm of b2](mult2) {\Large $\cdot$};
    \node[branch, above=of b2](b3) {};
    \node[blockplain, right=0.5cm of b3](tanh) {tanh};
    \node[above=of b3](b4) {};
    \node[blockplain, right=0.5cm of b4](s3){$\sigma$};
    \node[operator, right=1.5cm of b4](mult3) {\Large $\cdot$};
    \node[above=0.5cm of mult3](a1) {$a^{<t>}$}; 
    \node[blockplain, right=2.5cm of b4](tanh2) {tanh};
    
    \node[operator, right=4cm of b1](mult1) {\Large $\cdot$};
    \node[below=0.5cm of mult1](c0){$c^{<t-1>}$};
    \node[operator, right=4cm of b2](plus1) {$+$};
    \node[branch, above=2cm of  plus1](b5) {};
    \node[above=0.5cm of b5](c1) {$c^{<t>}$};

    \path[line] (a0)|-(s3);
    \path[line] (c0)--(mult1);
    \path[line] (b1)--(s1);
    \path[line] (s1)--node[above, pos=0.1]{$\Gamma_{f}$}(mult1);
    \path[line] (mult1)--(plus1);
    \path[line] (mult2)--(plus1);
    \path[line] (b2)--(s2);
    \path[line] (s2)--node[above, pos=0.25]{$\Gamma_{u}$}(mult2);
    \path[line] (b3)--(tanh);
    \path[line] (tanh)-|node[above, pos=0.5]{$\tilde{c}^{<t>}$}(mult2);
    \path[line] (b5)--(tanh2);
    \path[line] (tanh2)--(mult3);
    \path[line] (s3)--node[above]{$\Gamma_{o}$}(mult3);
    \path[line] (mult3)--(a1);
    \path[line] (plus1)--(c1);
\end{tikzpicture}

%% file: diag2.tex
\begin{tikzpicture}[auto]
    \node[block, fill=red!40](cosh) {Cosh};
    \node[state2, below=.5cm of cosh](relu3){ReLU\nodepart{two} Dense (1)};
    \node[state3, below=.5cm of relu3](relu2){Dropout\nodepart{two} ReLU\nodepart{three} Bi-LSTM 2 (1000)};
    \node[state3, below=.5cm of relu2](relu1){Dropout\nodepart{two}ReLU\nodepart{three} Bi-LSTM 1 (800)};
    
    \path[line] (relu3)--(cosh);
    \path[line] (relu2)--(relu3);
    \path[line] (relu1)--(relu2);
\end{tikzpicture}

%% file: diag1.tex
\begin{tikzpicture}[auto]
    \coordinate (c1) at (0,0);
    \node[below=1cm of c1](data) {\textbf{Data}};
    \node[block, right=.5cm of c1](techind) {
		\begin{minipage}{2.5cm}
			\begin{center}
                Feature Extraction\\
                \& Engineering\\
                851 in total
			\end{center}
		\end{minipage}};
    \node[block, right=.5cm of techind](feature) {
		\begin{minipage}{1.6cm}
			\begin{center}
                Feature\\ Selection\\
                RF Top 10
			\end{center}
		\end{minipage}};
    \node[block, right=.5cm of feature](split) {
		\begin{minipage}{1.5cm}
			\begin{center}
                Train/Test\\ Split
			\end{center}
		\end{minipage}};
    \node[block, right=.5cm of split](bilstm) {
		\begin{minipage}{1.5cm}
			\begin{center}
                Bi-LSTM\\
                3 Layers
			\end{center}
		\end{minipage}};
    \coordinate[right=.5cm of bilstm](c2);
    \node[below=of c2](pred) {\textbf{Prediction}};
    
    \path [line] (data) |- (techind);
    \path [line] (techind) -- (feature);
    \path [line] (feature) -- (split);
    \path [line] (split) -- (bilstm);
    \path [line] (bilstm) -| (pred);
\end{tikzpicture}